\documentclass[prl,twocolumn,showpacs,superscriptaddress,floatfix]{revtex4}
\usepackage{graphicx,amsfonts,amssymb,amsmath,hyperref}

\newif\ifhyper
\hypertrue
\ifhyper
\hypersetup{
   citecolor = {green},
   colorlinks = {true}, 
   urlcolor = {blue} 
}
\fi

\newcommand{\anb}{b^{\phantom\dagger}}
\newcommand{\crb}{b^\dagger}
\newcommand{\bn}{{\boldsymbol{n}}}
\newcommand{\bi}{{\boldsymbol{i}}}

\newcommand{\bk}{{\boldsymbol{k}}}

\newcommand{\rmi}{\mathrm i}

\begin{document}

\title{Creation and Manipulation of Anyons in the Kitaev Model}

\author{S\'ebastien Dusuel}
\email{sdusuel@gmail.com}
\affiliation{Lyc\'ee Louis Thuillier, 70 Boulevard de Saint Quentin,
  80098 Amiens Cedex 3, France}

\author{Kai Phillip Schmidt}
\email{schmidt@fkt.physik.uni-dortmund.de}
\affiliation{Lehrstuhl f\"ur theoretische Physik, Otto-Hahn-Stra\ss e 4, D-44221
Dortmund, Germany}

\author{Julien Vidal}
\email{vidal@lptmc.jussieu.fr}
\affiliation{Laboratoire de Physique Th\'eorique de la Mati\`ere Condens\'ee,
  CNRS UMR 7600, Universit\'e Pierre et Marie Curie, 4 Place Jussieu, 75252
  Paris Cedex 05, France}
    

\begin{abstract}
We analyze the effect of local spin operators in the Kitaev model on the
honeycomb lattice. We show, in perturbation around the isolated-dimer limit,
that they create Abelian anyons together with fermionic excitations which
are likely to play a role in experiments. We derive the explicit form of the
operators creating and moving Abelian anyons without creating fermions and show
that it involves multi-spin operations. Finally, the important experimental
constraints stemming from our results are discussed.
\end{abstract}


\maketitle

%
%
%
%
Anyons are particles which obey quantum statistics different from bosons
and fermions \cite{Wilczek82_1}. In condensed matter physics, such weird objects
are expected to arise, for instance, in fractional quantum Hall systems
\cite{Stormer83}, $p$-wave superconductors \cite{Read00}, or Josephson
junctions arrays \cite{Doucot04}, but have never been directly observed yet.
Within the last few years, anyons have also been shown to be of primer interest
for topological quantum computation \cite{Preskill_HP}.
Following Kitaev's seminal work related to this problematics \cite{Kitaev03}, several spin systems with
emerging Abelian and non-Abelians anyons have been proposed
\cite{Kitaev06,Yao07}, together with experimental proposals susceptible to
capture their fascinating braiding properties \cite{Duan03,Micheli06,Zhang07}.
In this perspective, Kitaev's honeycomb model \cite{Kitaev06} which only
involves two-spin interactions seems easier to realize than the toric code
model \cite{Kitaev03} based on four-spin interactions. 

As shown by Kitaev using 4th order perturbation theory, the honeycomb model can
be mapped onto the toric code, thus proving the existence of anyons in the model
\cite{Kitaev06}. However, orders {\em lower} than 4 also bring their physical
contributions and deserve a careful treatment, especially when considering
correlation functions and, more importantly, the action of spin operators.
The aim of this Letter is to shed light on these issues and to bring some
insights for experiments. We show that the original spin operators applied onto
the ground state create both anyons and fermions and we give the explicit
form of an operator creating anyons without fermions. 
At lowest nontrivial order, this operator involves a superposition of states involving one-spin and three-spin operations
on the ground state. 
To conclude, we discuss the constraints raised by our results for experiments in optical lattices.

%
%
\emph{The model ---}
%
%
Kitaev's honeycomb (or brickwall) model is a two-dimensional spin-$1/2$ system
described by the Hamiltonian
%
%
\begin{equation}
  \label{eq:ham}
  H=-\sum_{\alpha=x,y,z}\sum_{\alpha-\mathrm{links}}
  J_\alpha\,\sigma_i^\alpha\sigma_j^\alpha,
\end{equation} 
%
%
where the $\sigma_i^\alpha$'s are the usual Pauli matrices at site $i$. Without loss
of generality \cite{Kitaev06}, in the following, we assume $J_\alpha \geq 0$
for all $\alpha$ and $J_z\geq J_x,J_y$.

\begin{figure}[t]
  \includegraphics[width=\columnwidth]{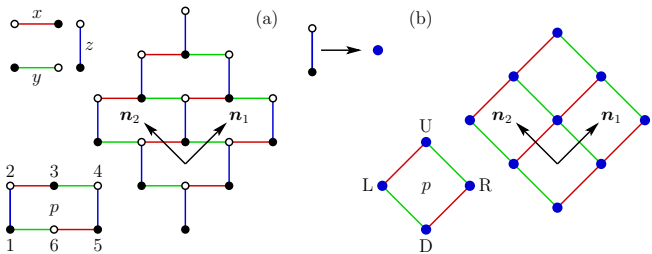}
  \caption{(color online).
    Mapping of the honeycomb (brick-wall) lattice (a) onto an effective square
    lattice (b) with unit basis vectors $\bn_1$ and $\bn_2$. The numbering of
    the sites of a plaquette $p$ is shown in both cases.}
  \label{fig:mapping_brickwall_square}
\end{figure}

For our purpose, it is convenient to map the spin model (\ref{eq:ham}) on the
honeycomb lattice onto an effective spin hardcore boson problem on a square
lattice. This mapping is achieved via the following rules
%
%
\begin{equation}
    \label{eq:mapping}
    \begin{array}{lcl}
    \sigma_{\bi,\bullet}^x=\tau_\bi^x (\crb_\bi+\anb_\bi)
    &,&
    \sigma_{\bi,\circ}^x=\crb_\bi+\anb_\bi ,\\ 
    \sigma_{\bi,\bullet}^y=\tau_\bi^y (\crb_\bi+\anb_\bi)
    &,&
    \sigma_{\bi,\circ}^y=\mathrm{i} \, \tau^z_\bi (\crb_\bi-\anb_\bi) , \\
    \sigma_{\bi,\bullet}^z=\tau_\bi^z
    &,&
    \sigma_{\bi,\circ}^z=\tau_\bi^z (1-2 \crb_\bi \anb_\bi),
    \end{array}
\end{equation}
%
%
where $\crb_\bi$ ($\anb_\bi$) is the creation (annihilation) operator of a
hardcore boson at site $\bi$ of the square lattice depicted in
Fig.~\ref{fig:mapping_brickwall_square} and $\tau_\bi^\alpha$ the Pauli matrices
of the effective spin at the same site. Here, we have attached each site of the
honeycomb lattice ($\bullet$ or $\circ$) to the $z$-dimer it belongs to.

In the following, we set $J_z=1/2$ so that the Hamiltonian reads 
%
%
\begin{equation}  \label{eq:ham_v2}
  H = -\frac{N}{2}+Q+T_0+T_{+2} + T_{-2},
 \end{equation}
%
%
where $N$ is the number of $z$-dimers,
%
%
\begin{eqnarray}
Q&=&\sum_\bi\crb_\bi\anb_\bi ,\\
 T_0&=&-\sum_\bi \left(J_x\, t_\bi^{\bi+\bn_1}+J_y\, t_\bi^{\bi+\bn_2}
    +\mathrm{h.c.}\right),\\
  T_{+2}&=&-\sum_\bi \left(J_x\, v_\bi^{\bi+\bn_1}+J_y\, v_\bi^{\bi+\bn_2}\right)
  =T_{-2}^\dagger,
\end{eqnarray}
%
%
with hopping and pair creation operators $t$ and $v$
%
%
\begin{eqnarray}
  \label{eq:t}
  t_\bi^{\bi+\bn_1}=\crb_{\bi+\bn_1}\anb_\bi\, \tau^x_{\bi+\bn_1},\quad
  t_\bi^{\bi+\bn_2}=-\rmi\, \crb_{\bi+\bn_2}\anb_\bi\,
  \tau^y_{\bi+\bn_2}\tau^z_\bi , &&\\
  \label{eq:v}
  v_\bi^{\bi+\bn_1}=\crb_{\bi+\bn_1}\crb_\bi\, \tau^x_{\bi+\bn_1},\quad
  v_\bi^{\bi+\bn_2}=\rmi\, \crb_{\bi+\bn_2}\crb_\bi\,
  \tau^y_{\bi+\bn_2}\tau^z_\bi.
\end{eqnarray}
%
%
We emphasize that this mapping is exact and simply relies on an alternative
interpretation of the four possible spin states  of a dimer (see
 \cite{Schmidt08,Vidal08_2} for details).
A key feature of this model is that $H$ commutes with  $W_p=\sigma_1^x\sigma_2^y\sigma_3^z\sigma_4^x\sigma_5^y\sigma_6^z
  =(-1)^{\crb_{\mbox{\tiny L}}\anb_{\mbox{\tiny L}}
    +\crb_{\mbox{\tiny D}}\anb_{\mbox{\tiny D}}}
  \tau_{\mbox{\tiny L}}^y\,\tau_{\mbox{\tiny U}}^z\,
  \tau_{\mbox{\tiny R}}^y\,\tau_{\mbox{\tiny D}}^z,
$
with notations given in Fig.~\ref{fig:mapping_brickwall_square}, so that it
describes a $\mathbb{Z}_2$ gauge theory. Further and more interestingly, it is
exactly solvable via several fermionization methods
\cite{Kitaev06,Feng07,Chen08}. Its phase diagram consists in a gapless phase
for $J_x+J_y>J_z$ and a gapped phase otherwise.
The latter contains high-energy fermions and low-energy Abelian anyons which are
the focus of the present work.
%
%
%
%

%
%
\emph{Perturbation theory ---}
%
%
Here, we focus on the isolated-dimer limit $(J_z \gg J_x, J_y)$ and, following
Kitaev, we perform a perturbative analysis of the gapped phase. A reliable tool
to investigate this problem is the continuous unitary transformations (CUTs)
method \cite{Wegner94}, used in its perturbative version
\cite{Stein97,Knetter00,Schmidt08}.
Within this approach, the Hamiltonian (\ref{eq:ham_v2}) is transformed into an
effective Hamiltonian $H_\mathrm{eff}$ which conserves the number of bosons
($[H_\mathrm{eff},Q]=0$). 
The cornerstone of the method is that $H_\mathrm{eff}$ is
{\em unitarily equivalent} to $H$, {\it i.~e.}, there exists a unitary transformation
$U$ such that $H_\mathrm{eff}=U^\dag H U$. This implies that $H$ and
$H_\mathrm{eff}$ {\em have the same spectrum but not the same eigenstates}. 
At order 1, the full effective Hamiltonian reads $H_\mathrm{eff}=-N/2+Q+T_0$,
and leads to high-energy fermionic excitations made of a hardcore boson and a
spin-string \cite{Schmidt08,Vidal08_2}.
At order 4, the effective Hamiltonian in the low-energy (no fermion) subspace is the toric code Hamiltonian \cite{Kitaev06}
%
%
\begin{equation}
  \label{eq:ham_toricode}
  \frac{H_\mathrm{eff}|_0}{N} = - \frac{1}{2}+\frac{J_x^2+J_y^2}{2}
  +\frac{J_x^4+J_y^4}{8}-\frac{J_x^2 J_y^2}{2N}\sum_{p}W_{p}.
\end{equation}
%
%

A simple way to obtain this result is to study the action of $H$, in perturbation,  on the low-energy eigenstates of the unperturbed Hamiltonian ($J_x=J_y=0$) which are ferromagnetic configurations of $z-$dimers. However, we emphasize that {\em the low-energy eigenstates of $H$ are not solely built from these states, but involve high-energy (antiferromagnetic) configurations, even at order 1} \cite{Vidal08_1}. This point
is the source of the discrepancy between the correlation functions computed in \cite{Zhang08_1} and exact results \cite{Baskaran07,Schmidt08}.
The perturbative expansion of the exact ground state spin-spin correlation functions at order 2 read
%
%
\begin{eqnarray}
\langle\sigma^x_{\bi,\circ}\sigma^x_{\bi+\bn_1,\bullet}\rangle=J_x &,& 
\langle\sigma^y_{\bi,\circ}\sigma^y_{\bi+\bn_2,\bullet}\rangle=J_y, \\
\langle\sigma^z_{\bi,\circ}\sigma^z_{\bi,\bullet}\rangle&=&1-(J_x^2+J_y^2).
\end{eqnarray}
%
%
These results clearly show that sticking to ferromagnetic dimer configurations
fails to capture the form of the low-energy states \cite{Vidal08_1,Zhang08_1}.

The Hamiltonian (\ref{eq:ham_toricode}), known as the toric code
\cite{Kitaev03}, has its ground state in the vortex-free sector
($W_p=+1, \forall p$) and contains Abelian anyons defined as plaquette
excitations ($W_p=-1$). The real challenge is obviously to observe these
anyons experimentally. In this perspective ultracold atoms in optical lattices
are good candidates \cite{Duan03} although devices using polar molecules have
also been proposed \cite{Micheli06}.
Recently, several detection protocols in optical lattices based on the
possibility to perform sequences of single-spin operations have been proposed
 \cite{Zhang07,Jiang08}. Unfortunately, although such operations create and move
anyons, they also create fermionic excitations which are susceptible to spoil
the detection process.
Our aim is to show that this crucial fact, always neglected in previous studies, is of major importance for experiments.

\begin{figure}[t]
  \includegraphics[width=\columnwidth]{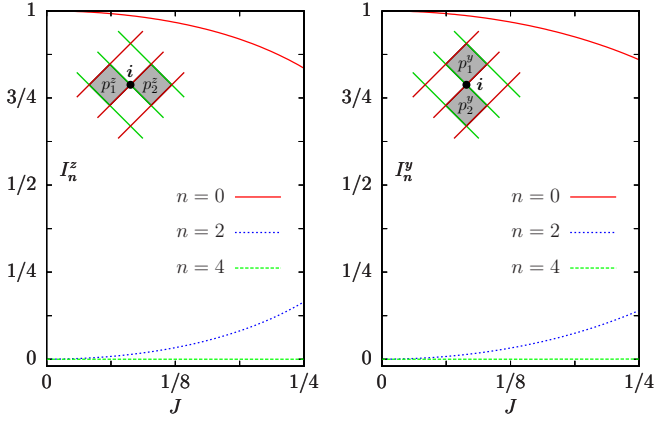}
  \caption{(color online).
    Behavior of the spectral weights $I_n^z$ (left) and $I_n^y$ (right)
    for fermion numbers $n=0,2,4$, as a function of the coupling $J=J_x=J_y$ for
    $J_z=1/2$. Grey plaquettes in the insets  $p_1^{z(y)}$ and
    $p_2^{z(y)}$ show where anyons are created under the action of
    $\tau_\bi^z$ (left) and $\tau_\bi^y$ (right).}
  \label{fig:Iyz}
\end{figure}

%
%
\emph{Single-spin operations ---}
%
%
Let us first discuss the action of single-spin operations on the ground state
and focus on $ \tau_\bi^z=\sigma_{\bi,\bullet}^z$. This operator is
used, in Refs.~\cite{Pachos07,Zhang07}, as the basic tile to create two anyons
on left and right plaquettes of the site $\bi$ [see inset in Fig.~\ref{fig:Iyz}
(left)].
As mentionned above the main problem is that it does not simply transform the
ground state $|0\rangle$ (belonging to the vortex-free sector) into the
lowest-energy state of the two-vortex configuration one seeks to obtain. To make
this statement quantitative, we consider the spectral weights
%
%
\begin{equation}
  \label{eq:weights}
  I^\alpha_n
  =\sum_{\bk} \left|\langle \{p_1^\alpha,p_2^\alpha\},n, \bk |
    \tau_\bi^ \alpha |0\rangle\right|^2,
\end{equation}
%
%
which obey the sum rule $\sum_n I^\alpha_n=1$.
Here, $|\{p\},n,\bk \rangle$ denotes the eigenstate of $H$ in a
sector given by an anyon configuration $W_p=-1$, and $n$ high-energy
quasiparticles with quantum numbers $\bk$. Plaquettes $p_1^\alpha$ and
$p_2^\alpha$ are neighbors of $\bi$ and depend on $\alpha=y,z$ (see insets of
Fig.~\ref{fig:Iyz}).
With these notations, the eigenstates of $H_\mathrm{eff}$ are given by
$U^\dag |\{p\},n,\bk \rangle$. To compute $I^z_n$, one is thus led to
determine how the observable $\tau_\bi^z$ is renormalized under the unitary
transformation $U$.
Within the CUT formalism, this can be achieved efficiently order by order in
perturbation \cite{Knetter03_2,Vidal08_2}.
At order 1, one gets
%
%
\begin{equation}
  \label{eq:rensz1}
  U^\dagger \tau_\bi^z  U = \tau_\bi^z \left[1+
    \left(J_x v_{\bi-\bn_1}^{\bi} +J_y v_{\bi-\bn_2}^{\bi}
      +\mbox{ h.c.}\right)\right],
\end{equation}
%
%
which clearly shows that pairs of quasiparticles are created or annihilated. In
addition, one can see that the action of $\tau_\bi^z$ on the ground state
yields a superposition of states with the appropriate two-vortex configuration,
but with different number of quasiparticles. 
We computed the renormalization of $\tau_\bi^z$ up to order 6 but we emphasize
that quasiparticles arise at order 1 which makes this phenomenon strongly
relevant.
The perturbative expansion of $U^\dagger \tau_\bi^z U$ allows a straighforward
calculation of the overlap probability which reads at order 6: $I^z_0=1-I^z_2 $ with
%
%
\begin{eqnarray}
 I^z_2&=&\left(J_x^2+J_y^2 \right)+\frac{3}{2}\left(J_x^4+J_y^4\right)+ 4 J_x^2 J_y^2+\nonumber \\
 && \frac{7}{2} \left(J_x^6+J_y^6 \right)+ \frac{43}{2} \left(J_x^2 J_y^4+J_x^4 J_y^2 \right).
\end{eqnarray}
%
%
Note that only even $n$ contributions are nonvanishing. In addition, we found
that $I^z_{n\geq 4}=0$ at order 6, showing the sum rule is fulfilled.

%
%
\emph{Multiple-spin operations ---}
%
%
The same approach can be used to analyze the action of
$\tau_\bi^y=\sigma_{\bi,\bullet}^y\sigma_{\bi,\circ}^x$ on the ground state,
which creates two anyons on the up and down plaquettes of the site $\bi$ [see
inset in Fig.~\ref{fig:Iyz} (right)] as well as quasiparticles. As previously,
we computed the spectral weights up to order 6 and obtained: $I^y_0=1-I^y_2-I^y_4$ with
%
%
\begin{eqnarray}
I^y_2&=&\left(J_x^2+J_y^2 \right)+\frac{3}{2}\left(J_x^4+ J_y^4\right)+ \nonumber \\
 &&  \frac{7}{2} \left(J_x^6+J_y^6 \right)+ 3 \left(J_x^2 J_y^4+J_x^4 J_y^2 \right), \\
I^y_4&=& J_x^2 J_y^4+J_x^4 J_y^2.
\end{eqnarray}
%
%
The behavior of $I^z_n$ and $I^y_n$ are plotted in Fig.~\ref{fig:Iyz} as
a function of the parameter $J_x=J_y=J$. One can directly observe the growing
weight (for increasing $J$) of the excited states after the operations $\tau_z$
or $\tau_y$ have been applied on the ground state.
%
%
%
%

\begin{figure}[t]
  \includegraphics[width=\columnwidth]{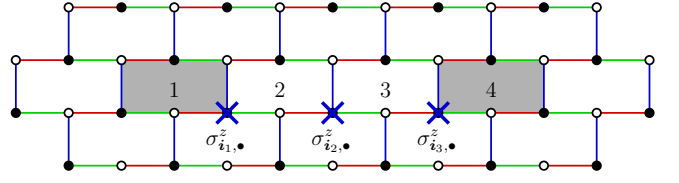}
  \caption{(color online).
Action, in the vortex-free sector, of the string operator $S=\sigma_{\bi_3,\bullet}^z \sigma_{\bi_2,\bullet}^z \sigma_{\bi_1,\bullet}^z $. Each operator flips the two plaquettes adjacent to the $z$-dimer it is attached to but also creates fermionic excitations (not shown).}
  \label{fig:braid_z}
\end{figure}

Other important quantities are string operators which are crucial in
experiments to perform braiding of anyons. To investigate the effect of such
multiple-spin operations, let us consider a string
$S=\prod_{a=1,m} \sigma_{\bi_a,\bullet}^z$ along a horizontal line of the
original brickwall-lattice (see Fig.~\ref{fig:braid_z} with $m=3$ for
notations). 
When such a sequence is applied on the ground state $|0\rangle$, the
probability to find the final state in the lowest-energy state with anyons at
plaquettes 1 and $(m+1)$ is given, at order 2, by
%
%
\begin{equation}
  \label{eq:proba}
  \left|\langle \{1,m+1\},0 | S |0\rangle\right|^2=1-m(J_x^2+J_y^2),
\end{equation}
%
%
which obviously coincides with $I^z_0$ for $m=1$. Note that we dropped the
$\bk$ index which is useless for states without fermions.
This result shows that, at this order, the role played by the excited states is
directly proportional to the length of the string.
%
%
%
%

%
%
\emph{Anyons without fermions ---}
%
%
At this stage, we have shown that the action of the original spins
$\sigma_i^\alpha$ on the ground state does generate unwanted excited states as
can be seen at lowest nontrivial order in the perturbation theory.
One may wonder how to create anyons without fermions.
In other words, we wish to determine the operators that only flip two $W_p$'s
and nothing else. 

As an example, we focus on the operator creating two
vortices at the left and right plaquettes of a given site $\bi$ [see inset
Fig.~\ref{fig:Iyz} (left)].
Let us denote this operator $\Omega_\bi$, and show how to compute its
perturbative expansion, $\Omega_\bi=\sum_{k\in \mathbb{N}} \Omega_\bi^{(k)}$,
where $\Omega_\bi^{(k)}$ contains all operators of order $k$ and thus associated
to $J_x^l J_y^m$ (with $l+m=k$).
The aim is to obtain
${\Omega_\bi}_\mathrm{eff}=U^\dagger\Omega_\bi U=\tau_\bi^z$ which indeed leads
to $I_0^z=1$.
At order 0, operators are not renormalized so that one obviously has
$\Omega_\bi^{(0)}=\tau_\bi^z$.
The renormalization of $\Omega_\bi$ under the unitary transformation $U$ reads
%
%
\begin{eqnarray}
  \label{eq:expansion_Omega_eff}
  {\Omega_\bi}_\mathrm{eff}=\sum_{k\in \mathbb{N}}
  {\Omega_\bi}_\mathrm{eff}^{(k)}
  =\sum_{k\in \mathbb{N}} U^\dag \Omega_\bi^{(k)} U
  =\sum_{k \in \mathbb{N}}\sum_{l\in \mathbb{N}}
  {{\Omega_\bi}_\mathrm{eff}^{(k),[l]}},
\end{eqnarray}
%
%
where ${{\Omega_\bi}_\mathrm{eff}^{(k),[l]}}$ is of order $(k+l)$. Order $0$
being already fulfilled, let us write this constraint at order $1$, using
Eq.~(\ref{eq:rensz1}) and ${{\Omega_\bi}_\mathrm{eff}^{(k),[0]}}=\Omega_\bi^{(k)}$
%
%
\begin{equation}
  \label{eq:expansion_Omega1}
  0={{\Omega_\bi}_\mathrm{eff}^{(0),[1]}}
  +{{\Omega_\bi}_\mathrm{eff}^{(1),[0]}}=\left(J_x v_{\bi-\bn_1}^{\bi} +J_y v_{\bi-\bn_2}^{\bi}
    +\mbox{h.c.}\right)+\Omega_\bi^{(1)}.
\end{equation}
%
%
This equation yields $\Omega_\bi^{(1)}$, so that using the inverse mapping of
Eq.~(\ref{eq:mapping}) one finally gets, in the original spin language and at
order 1
%
%
\begin{eqnarray}
    \label{eq:nofermion1}
    \Omega_\bi&=&\sigma_{\bi,\bullet}^z+\frac{1}{2}\Big[\\
    & &
    J_x
    \Big(
    \sigma_{\bi-\bn_1,\bullet}^z\sigma_{\bi-\bn_1,\circ}^y
    \sigma_{\bi,\bullet}^y 
    -\sigma_{\bi-\bn_1,\circ}^x\sigma_{\bi,\bullet}^x\sigma_{\bi,\circ}^z
    \Big)+ \nonumber \\
    & &
    J_y
    \Big(
    \sigma_{\bi-\bn_2,\bullet}^z\sigma_{\bi-\bn_2,\circ}^x
    \sigma_{\bi,\bullet}^x 
    -\sigma_{\bi-\bn_2,\circ}^y\sigma_{\bi,\bullet}^y\sigma_{\bi,\circ}^z
     \Big) 
    \Big].\nonumber
 \end{eqnarray}
%
%

This expression shows that to create anyons without fermions, one must be able
to build a delicate superposition of single spin-flip and three spin-flip states. 
Furthermore, the weights of the latter have to be fine-tuned since they depend on the precise values of the couplings which may be experimentally challenging. Note also that higher order corrections would even involve more spins.

%
%
\emph{Experimental discussion ---}
%
%
Let us now put our results in an experimental perspective, setting, for
simplicity, $J_x=J_y=J$ and $\gamma=J/J_z$.
Equation (\ref{eq:proba}) shows that the repeated action of $m$ spin-flips
decreases the weight of the (pure) two-anyon state in the low-energy subspace. One should thus
work with small enough values of $\gamma$.
For a reasonable number $m\simeq 25$ of operations needed to perform a
braiding of anyons, and assuming that $m \gamma^2/2\simeq 1/2$
\cite{Note} leads to conclusive experiments, 
one is led to choose $\gamma\lesssim 0.2$.

From a practical point of view, we also wish to mention that detecting anyons
using ground-state two-spin correlation functions after an anyonic braiding is
rather difficult.
Indeed, contrary to what is mentioned in \cite{Zhang07}, these functions do not
change sign in the presence of anyons. The latter are only responsible for 4th
order corrections in $\langle\sigma_{\bi,\circ}^z\sigma_{\bi,\bullet}^z\rangle$
whose leading term is of order 0 \cite{Schmidt08}, and for 3rd order changes in
$\langle\sigma_{\bi,\circ}^x\sigma_{\bi+\bn_1,\bullet}^x\rangle$ and
$\langle\sigma_{\bi,\circ}^y\sigma_{\bi+\bn_2,\bullet}^y\rangle$ whose leading
terms are of order 1 \cite{Vidal08_2}.
In this respect, the more efficient and sophisticated set-up recently proposed in \cite{Jiang08} 
is an interesting alternative.

We end this section with some remarks about temperature issues
in experiments, based on the orders of magnitude $J_z/h=5\mathrm{kHz}$ and
$J/h=1\mathrm{kHz}$ \cite{Trotzky08} ($\gamma=0.2$).
To perform anyonic interferometry experiments \cite{Zhang07,Jiang08}, the
temperature of the system should be small enough to prevent thermal excitations
of unwanted anyons.
Assuming thermal equilibrium, the typical low-energy scale
$J_\mathrm{eff}= J_z \gamma^4 /16$ appearing in front of the plaquette term in
Eq.~(\ref{eq:ham_toricode}) leads to the constraint
$T\ll J_\mathrm{eff} /k_\mathrm{B}\simeq 20\mathrm{pK}$.
Such a temperature is lower than recently reached temperatures
\cite{Leanhardt03}, but this problem may be circumvented by working out of
equilibrium.

To sum up, we have shown that unless experimentalists manage to apply
the operators given in Eq.~(\ref{eq:nofermion1}), they will have to make
sure the constraint $\gamma=J/J_z\lesssim 0.2$ is fulfilled. This constraint,
arising from the renormalization of spin operators at order 1 in perturbation,
could lead to a problematic upper bound on the temperature. The latter indeed
scales as $\gamma^4$ since the toric code Hamiltonian arises as an effective
Hamiltonian of Kitaev's honeycomb model only at order 4 in perturbation.
Finally, let us mention that the time evolution of the system between spin operations may also lead to an unwanted dynamics of the fermions but this difficulty can be circumvented by freezing the system via an increase of the optical lattice barriers.
\acknowledgments

We are indebted to Liang Jiang for fruitful discussions and for his help concerning experiments. 
KPS acknowledges ESF and EuroHorcs for funding through his EURYI.


\begin{thebibliography}{27}
\bibitem[{Not()}]{Note}
\bibinfo{note}{Remember that to recover the physical dimensions one has to
  replace $J$ by $J/(2J_z)$ in all formulae.}
\end{thebibliography}

\end{document}